\documentstyle[11pt,newpasp,twoside,epsf]{article}
\def\edcomment#1{\iffalse\marginpar{\raggedright\sl#1\/}\else\relax\fi}
\reversemarginpar

\begin{document}
\title{Milliarcsecond structure of methanol masers in L1206 and GL2789}
 \author{M.A. Voronkov and V.I. Slysh}
\affil{Astro Space Center, Profsouznaya st. 84/32, 117997 Moscow, Russia}

\begin{abstract}
We report results of EVN interferometric study of two star-forming regions
L1206 and GL2789 in the brightest methanol maser
line at 6.7~GHz. Using measured absolute
positions both methanol masers were identified with protostars which
are sources of bipolar outflows. Both masers consist of several maser spots,
with some of them being aligned in a linear structure with velocity gradient
probably delineating edge-on circumstellar disks.
We estimated the radii of such disks to be 140 AU and 280 AU (or 700 AU for
the whole structure treated as a disk) for L1206 and GL2789 respectively.
The brightness temperatures of the most intense fe\-atu\-res in L1206 and GL2789
are at least $1.1\times10^{10}$~K and $1.4\times10^9$K respectively.
\end{abstract}

\section{Introduction}
The $5_1-6_0$~A$^+$ methanol transition at 6.7~GHz produces the
brightest known methanol masers. Many recent interferometric studies of such
masers reveal geometrically ordered structures formed by maser spots
sometimes with velocity gradients (Norris et al. 1993;
Phillips et al. 1998; Minier et al. 2001). Such linear structures can
be explained in the model of the rotating Keplerian disk seen edge-on.
In this paper we present results of EVN observations of two methanol
masers L1206 and GL2789, which were previously detected in Medicina survey
(Slysh et al. 1999). These masers are not associated with 
ultra-compact {\sc H ii} regions, in contrast to the majority of strong
masers.
\begin{figure}[!hbt]
\plottwo{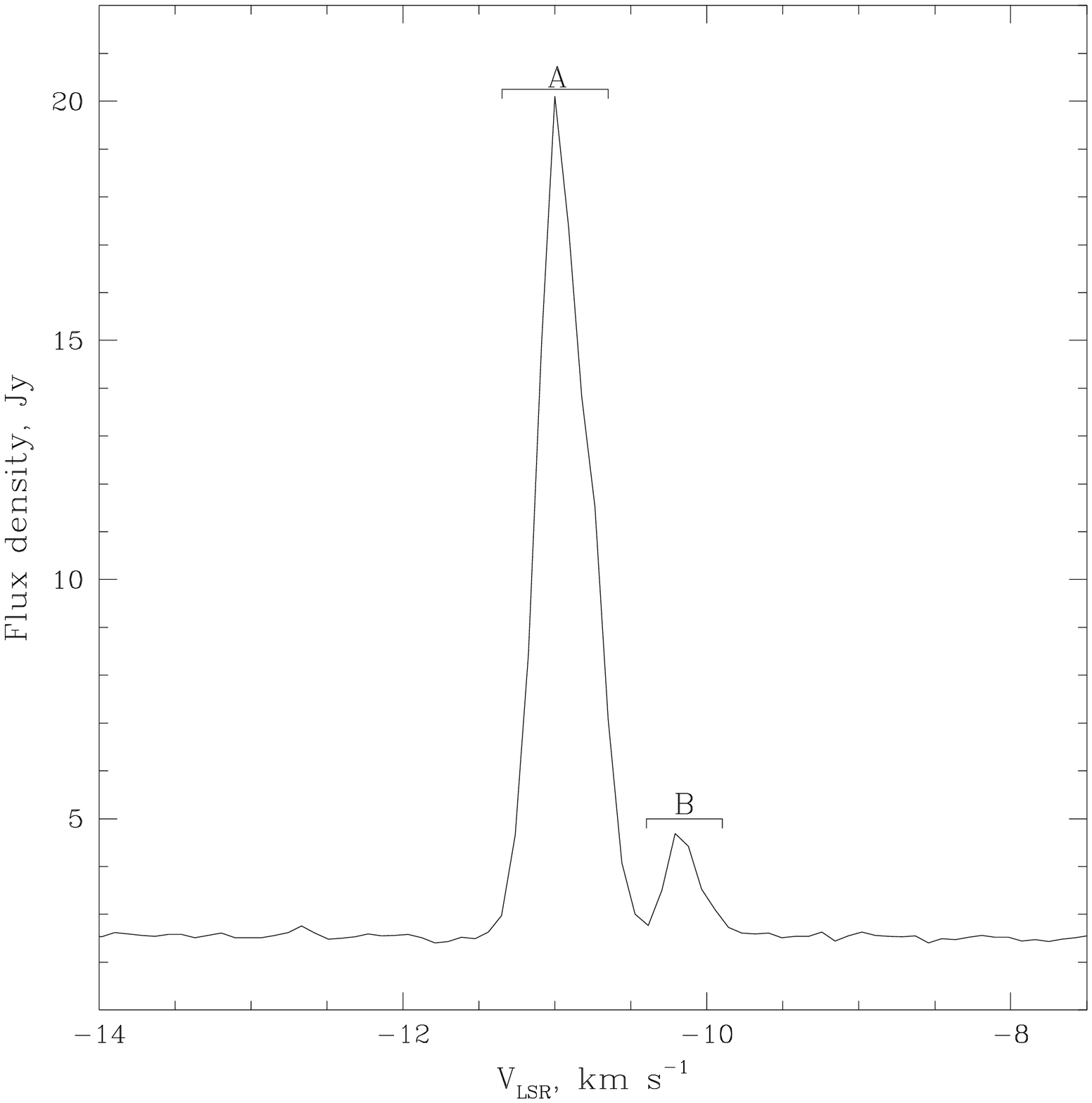}{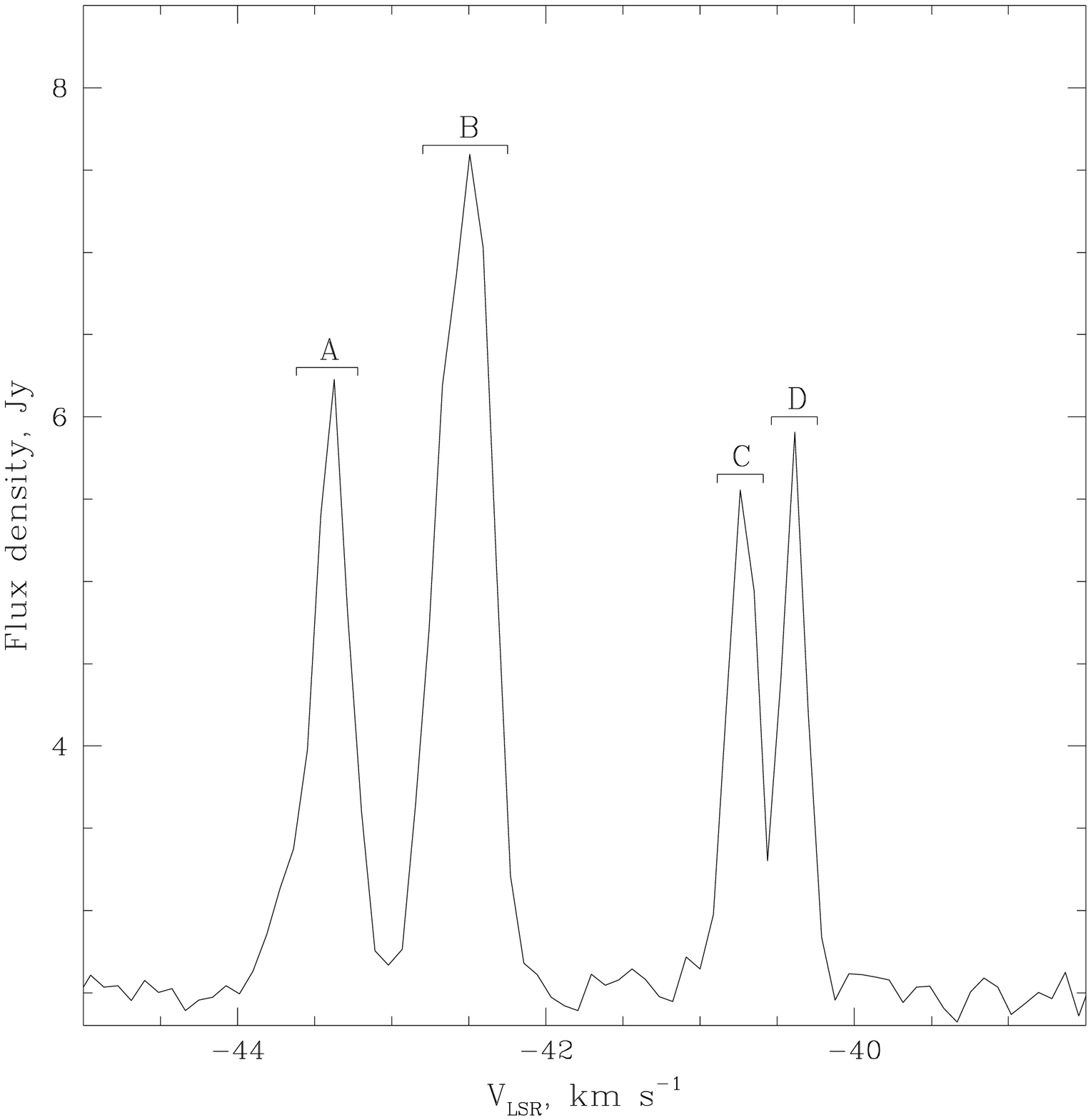}
\caption{The correlated spectra on the Effelsberg $-$ Medicina
baseline for methanol masers in L1206 (left) and GL2789 (right).}
\end{figure}
\begin{figure}[!hbt]
\plottwo{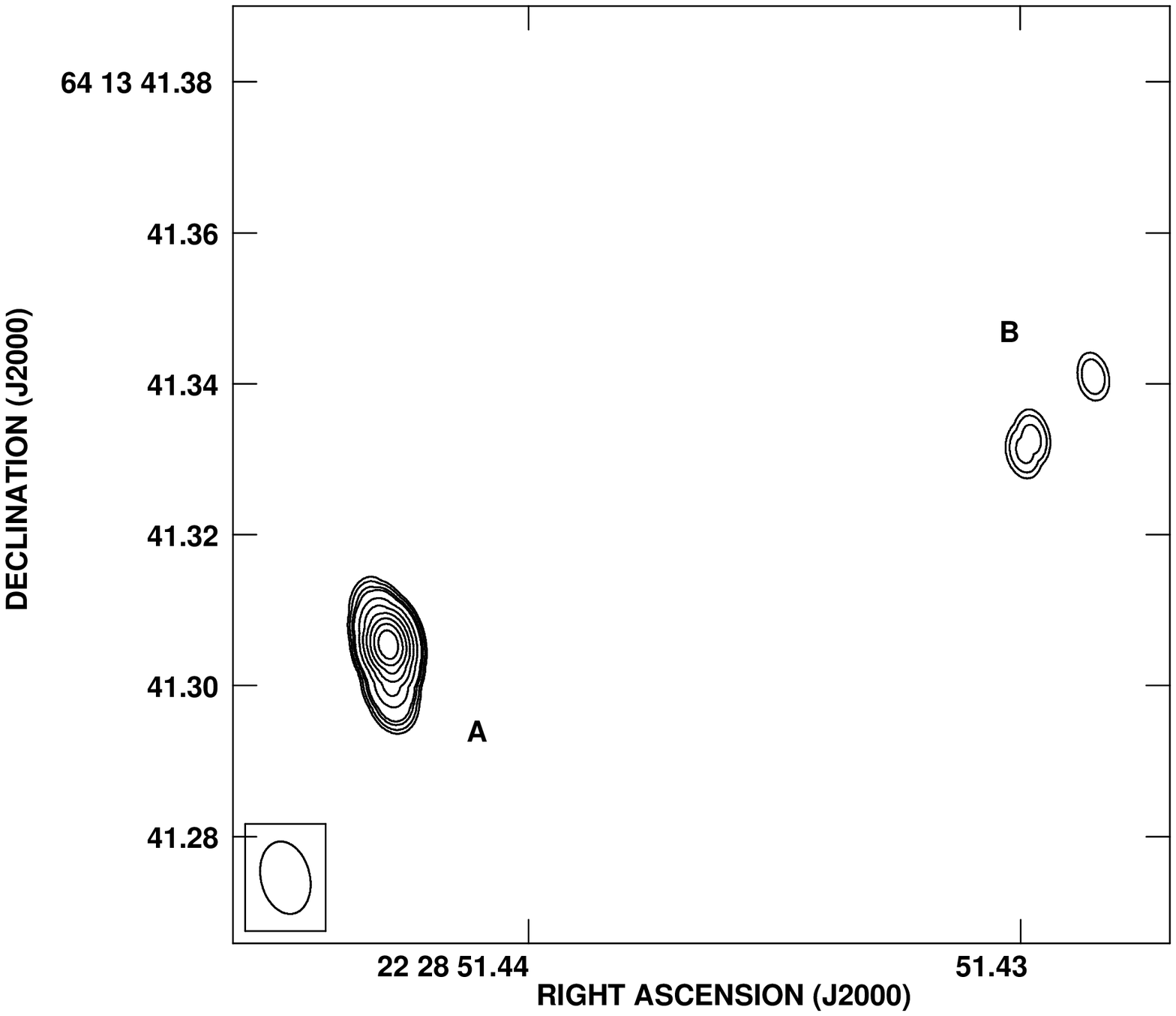}{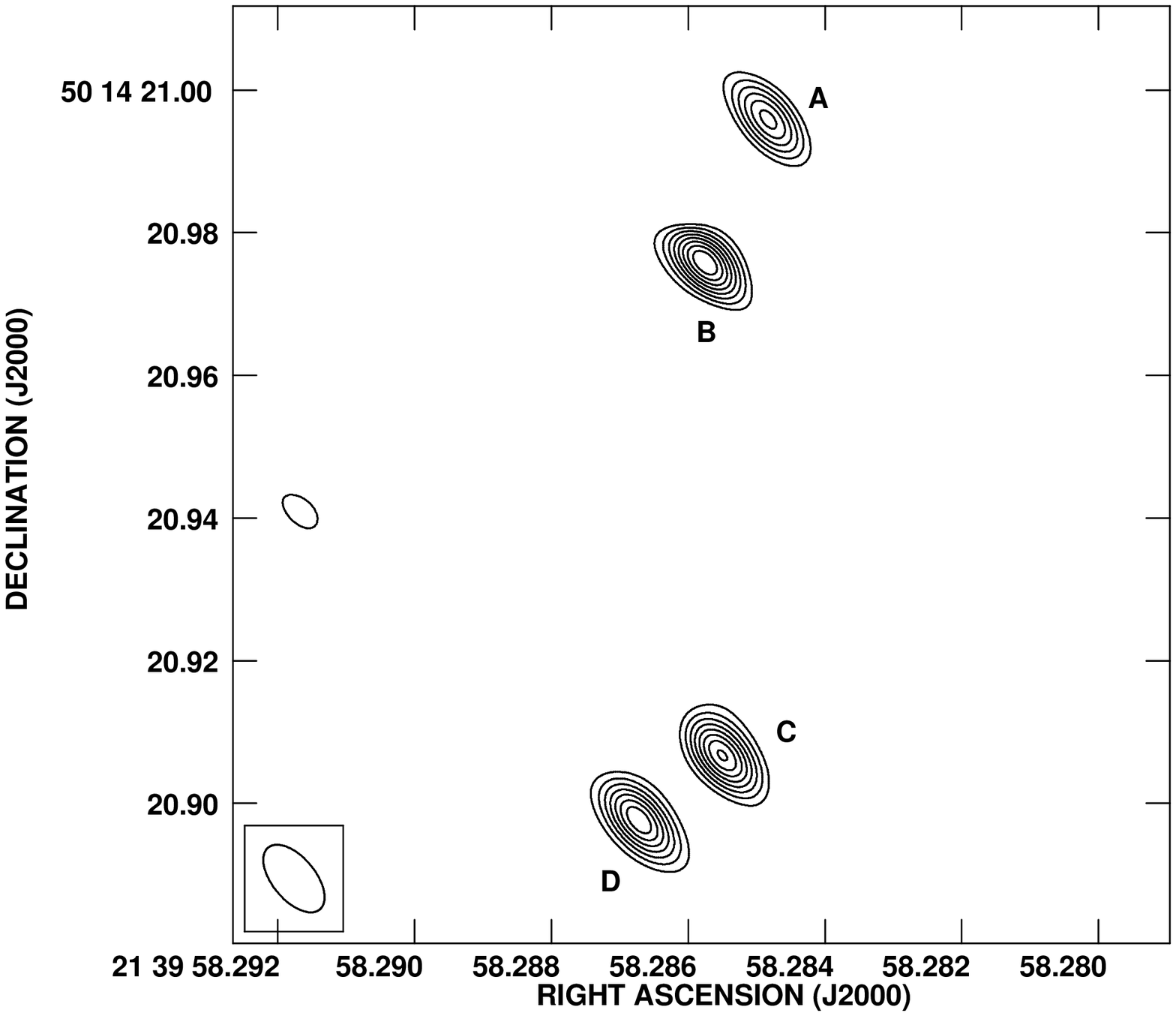}
\caption{The component map of the masers. Left is L1206, contours are
0.6$\times$(1, 1.5, 2, 2.3, 3, 4, 5, 6, 7, 8, 9)~Jy/beam. Right is
GL2789, contours are 0.22$\times$(2,3,4,5,6,7,8,9)~Jy/beam. 
All spectral channels are summed together.}
\end{figure}

\section{Results}
The correlated spectra of L1206 and GL2789 on the short baseline
Effelsberg $-$ Medicina are shown in Fig.~1. There are 2
and 4 features for L1206 and GL2789 spectrum respectively, maser spots
corresponding to each spectral feature are shown in the
component map in Fig.~2. The most intense L1206 feature A
at $-$11 km~s$^{-1}$ and GL2789 feature B at $-$42.5 km~s$^{-1}$ were
taken as reference features during self-calibration. The absolute positions
of these features were determined using the fringe rate method yielding
$\alpha=22^h28^m51\fs44\pm0\fs02$,
$\delta=64\deg13\arcmin41\farcs31\pm0\farcs1$ (J2000.0) for L1206
feature~A and 
$\alpha=21^h39^m58\fs29\pm0\fs01$,
$\delta=50\deg14\arcmin21\farcs0\pm0\farcs1$ (J2000.0) for GL2789
feature~B. Such positions allow us to identify both masers with
protostellar sources seen in the infrared, with the maser in GL2789 being
projected directly onto the center of spherical object N0 (Ressler \& Shure
1991; Minchin et al. 1991). The sizes of feature A in L1206 and
feature B in GL2789 have been measured to be less than 5.4~mas
and 6.5~mas, implying brightness temperatures more than
$1.1\times10^{10}$~K and $1.4\times10^9$~K, respectively. These high values can
be explained by the current maser models by Sobolev et al. (1997).

\section{Discussion}
\subsection{L1206}
\begin{figure}
\plottwo{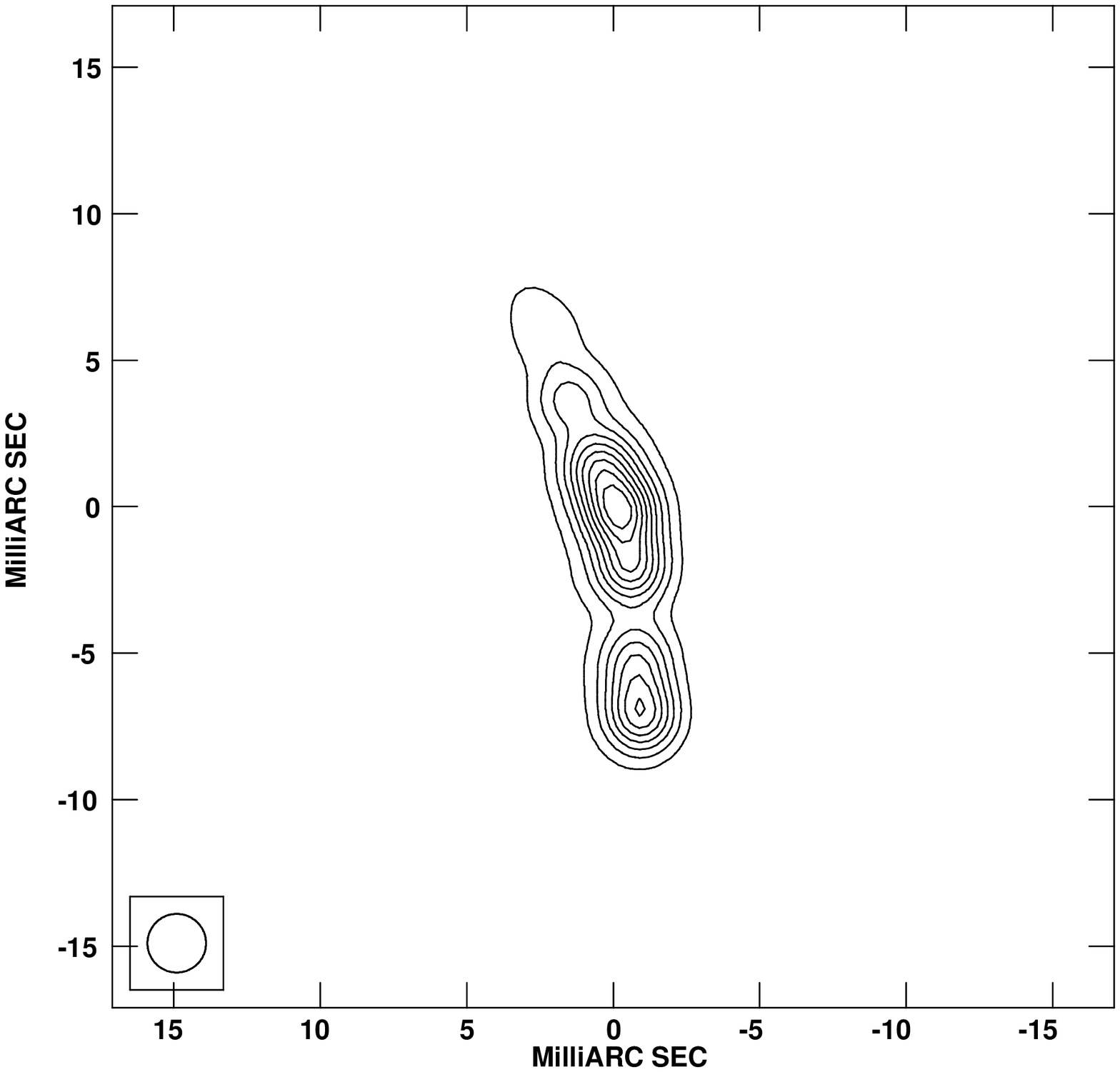}{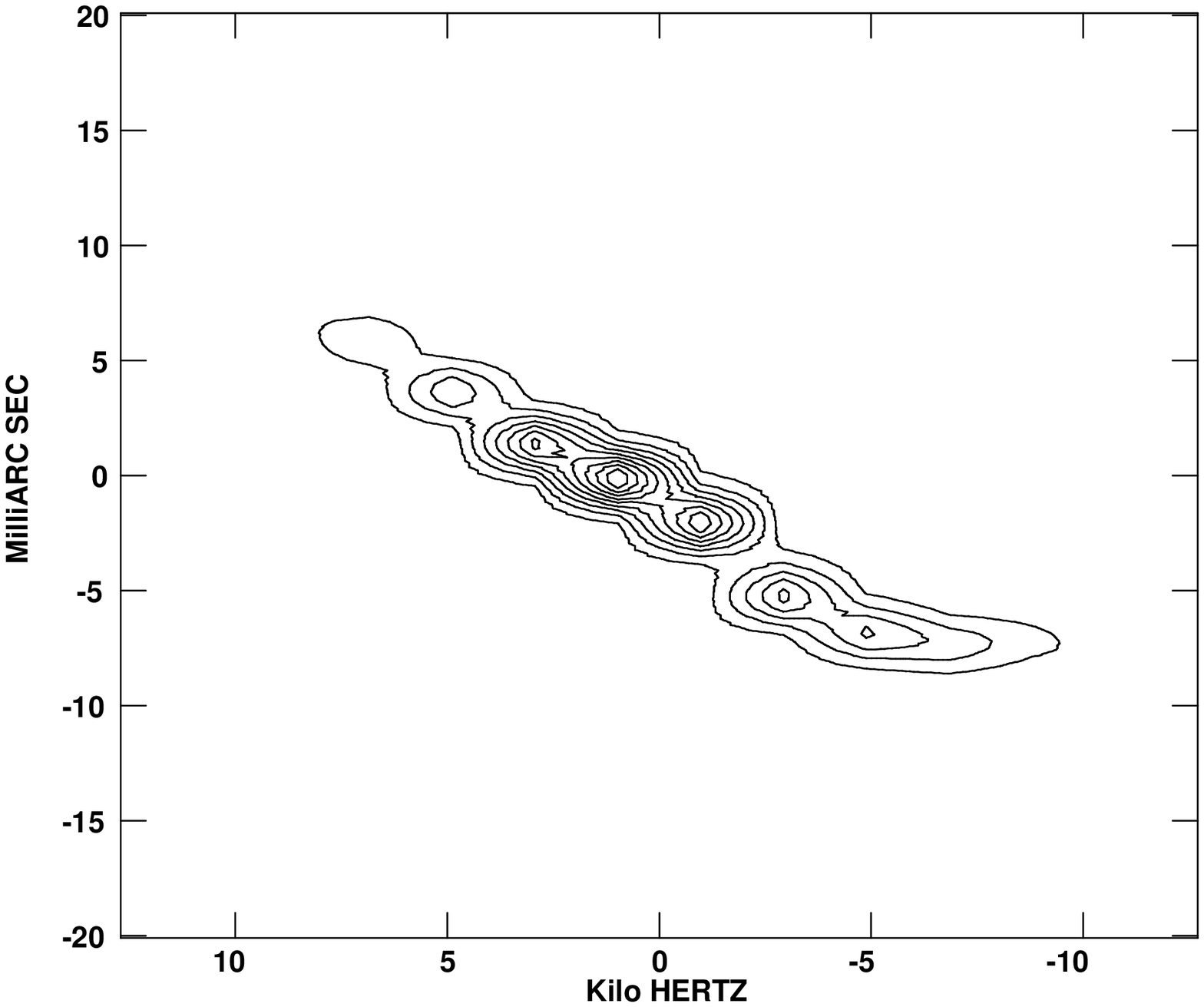}
\caption{Left: a super-resolution image of the L1206 feature A, all spectral
channels are averaged together.
Contours are 0.13$\times$(1, 2, 3, 4, 5, 6, 7, 8, 9)~Jy/beam.
Right: declination -- frequency diagram for the left image.
Contours are
0.43$\times$(1, 2, 3, 4, 5, 6, 7, 8, 9)~Jy/beam.
}
\end{figure}
The maser consists of two maser spots separated by about 200 mas, or
200~AU at the distance of 1~kpc. The most intense feature A in L1206
has, in turn, its own structure: individual spectral channels inside the
line profile produce unresolved images with our resolution, but images
corresponding to different spectral channels have slightly different
positions. To investigate this structure the feature A was
imaged using the super-resolution technique with the circular 2~mas
restoring beam, as compared to the synthesized beam of 9.7$\times$6.5~mas
for L1206 and 11.3$\times$5.6~mas for GL2789. The result of this mapping
shown in the left side of Fig.~3
is a structure which is very close to the line in the south-north direction.
There is a linear velocity gradient along this structure shown
at declination -- frequency diagram
in the right side of Fig.~3. The least square fit to the diagram gives a
slope $\Delta V_{km~s^{-1}}/\Delta\delta_{mas}=-(0.040\pm0.006)$~km~s$^{-1}$~mas$^{-1}$.
Assuming that maser spots delineate a disk with Keplerian velocity field,
one can estimate the ratio $M/R^3$ for such disk, where M~is the star mass
and R~is the disk radius. Note, that this ratio can be determined
with considerably higher accuracy than the mass or the radius alone. 
Assuming that the disk is not inclined and the distance to the
source is 1~kpc,
one obtains $M/R^3=(1.8\pm0.3)\times10^{-6}$~M$_\odot$AU$^{-3}$.
This value is an order of magnitude higher than
that obtained by Minier et al. (2001) for other sources.
The disk radius delineating by the small scale structure
of the feature A in L1206 is about 140~AU, if we assume that the central
object has a mass of 5~M$_\odot$ as follows from its infrared luminosity
(Ressler \& Shure 1991).  The part of this disk traced by methanol masers
(the total size of the spot A) is only 15~AU in extent. 
Linear distance between features A and B is 
comparable to the estimated disk size. However, the direction from the
feature A to
the feature B is not coincident with the direction traced by the 
disk and is close to the direction perpendicular to the disk plane.
Also, this direction is close to the direction of the outflow seen in the
map of
Ressler \& Shure (1991). Hence, it is possible that the feature A
delineates a disk around the protostellar object and the feature B may
originate in the outflow.
\subsection{GL2789}
The map of GL2789 consists of 4 maser spots scattered mostly in the
south-north direction. Assuming the kinematic distance to the source
of about 6~kpc, the linear extent of the global structure will be
600~AU. The most intense feature B probably has a small scale structure
similar to that of L1206. By analogy to L1206, one obtains
$M/R^3=(5\pm1)\times10^{-7}$~M$_\odot$AU$^{-3}$. This value is also higher than
that obtained by Minier et al. (2001). Assuming that the mass of the central
object is 10~M$_\odot$, which is following from infrared luminosity
(Minchin et al. 1991, and references therein), the disk radius will be
about 280~AU. For the whole structure seen in Fig.~2, we obtained
$M/R^3=(2.6\pm0.3)\times10^{-8}$~M$_\odot$AU$^{-3}$, the value
being in agreement with Minier et al. (2001) data. It implies the disk
radius of about 700~AU with the above mentioned mass of the central source.

\section{Conclusions}
Both masers were identified with protostars seen in the infrared.
Brightness temperatures of the most intense features exceed
$1.1\times10^{10}$~K  and  $1.4\times10^9$~K for L1206 and GL2789
respectively.
The L1206 feature A and probably GL2789 feature B are geometrically ordered
structures with velocity gradients, which may represent edge-on disks.
The size of such disk for the feature A of L1206 is estimated to be
about 140 AU, and for GL2789 feature B to be about 280 AU. For the
large scale GL2789 structure the estimated disk radius is about 700 AU.
The feature A in L1206 is probably originated in the disk and the
feature B may be connected with an outflow seen in the infrared.
\acknowledgements We would like to thank Dr.~F.~Palagi for valuable discussions,
which undoubtedly improved the content of this paper. This work was supported in
part by INTAS (grant no. 97-1451), RFBR (grant no. 01-02-16902) and
Radio Astronomy Research and Education Center (project no. 315).


\begin{references}

\reference Minchin N. R., Hough J. H., Burton M. G., Yamashita T.
1991, \mnras, 251, 522

\reference Minier V., Conway R. S., Booth R. S. 2001, \aap, 362, 1093

\reference Norris R. P., Whiteoak J. B., Caswell J. L., Wieringa M. H.,
Gough R. G. 1993, \apj, 412, 222

\reference Phillips C. J., Norris R. P., Ellingsen S. P.,
McCulloch P. M. 1998, \mnras, 300, 1131

\reference Ressler M. E., Shure M. 1991, \aj, 102, 1398

\reference Slysh V. I., Val'tts I. E., Kalenskii S. V., Voronkov M. A.,
   Palagi F., Tofani G., Catarzi M. 1999, \aaps, 134, 115

\reference Sobolev A. M., Cragg D. M., Godfrey P. D. 1997,
\mnras, 288, L39

\end{references}
\end{document}